\documentstyle[12pt]{article}

\newtheorem{theorem}{Theorem}
\newtheorem{definition}{Definition}

\newcommand{\p}{\partial}
\newcommand{\om}{\omega}
\newcommand{\e}{\epsilon}

\def\rit{I\!\!R}

\begin{document}

\title{Dissipative Quasigeostrophic Motion under Temporally
Almost Periodic Forcing
\footnote{This work was partly supported by the
National Science Foundation Grant DMS-9704345 and
the DFG Forschungschwerpunkt
``Ergodentheorie, Analysis und effiziente Simulation
dynamischer Systeme". This work was begun during a
visit to the Oberwolfach  Mathematical  Research
Institute, Germany.}  }

\author{
Jinqiao Duan\\
\vspace{-2mm}
\small\sl Department of Mathematical Sciences,\\
\vspace{-2mm}
\small\sl Clemson University, Clemson, South Carolina 29634, USA.\\
\vspace{-2mm}
\small\sl E-mail: duan@math.clemson.edu\\
\vspace{-5mm}
\and \protect{\newline}
\vspace{-1mm}
Peter E.~Kloeden \\
\vspace{-2mm}
\small\sl FB Mathematik, Johann Wolfgang Goethe Universit\"at,\\
\vspace{-2mm}
\small\sl  D-60054 Frankfurt am Main, Germany\\
\vspace{-2mm}
\small\sl E-mail: kloeden@math.uni-frankfurt.de
}

\vspace{-2mm}

%\date{December 25, 1998}

\maketitle

\begin{abstract}

The full nonlinear dissipative quasigeostrophic model is shown to have a unique
temporally almost periodic solution when the wind forcing is temporally almost
periodic under suitable constraints on the spatial square--integral of the wind
forcing and the $\beta$ parameter, Ekman number, viscosity and the domain size.
The proof involves the pullback attractor for the associated nonautonomous
dynamical system.

\bigskip


\end{abstract}

\bigskip

{\bf Short running title:} Almost Periodic Quasigeostrophic Motion

\newpage

\section{Introduction}

The barotropic quasigeostrophic (QG) flow model is derived as an approximation
of the rotating shallow water equations by an  asymptotic expansion in a small
Rossby number. The lowest order approximation, which is also the
conservation law
for the $0$th order potential vorticity,  gives the barotropic QG equation
(\cite{Pedlosky}, page 234)
\begin{eqnarray} \label{qg}
\Delta \psi_t + J(\psi, \Delta \psi) + \beta \psi_x	=  \nu \Delta^2
\psi - r
\Delta \psi + f(x,y,t),
\end{eqnarray}
where $\psi(x,y,t)$ is the stream function, $\beta$ $>$ $0$ the meridional
gradient
of the Coriolis parameter,  $\nu$ $>$ $0$  the viscous dissipation constant,
 $r$ $>$ $0$ the Ekman dissipation constant and $f(x,y,t)$ the wind forcing.
In addition, $\Delta$ $=$ $\p_{xx}+\p_{yy} $ is the Laplacian operator in the
plane and $J(f,g)$ $=$ $f_xg_y -f_yg_x$ is the Jacobian operator.

Equation (\ref{qg}) can be rewritten in terms of  the relative vorticity
$\om (x,y,t)$
$=$ $\Delta \psi(x, y,t)$ as
\begin{equation}  \label{eqn}
 \om_t + J(\psi, \om ) + \beta \psi_x =\nu \Delta \om - r \om  +  f(x,y,t) \; ,
\end{equation}
For an arbitrary bounded planar domain $D$ with area $|D|$ and piecewise
smooth
boundary this equation  can be supplemented by  homogeneous Dirichlet boundary
conditions for both $\psi$ and $\om = \Delta \psi$,
namely, the no-normal flow and slip  boundary conditions
(\cite{Pedlosky2}, page 34)
\begin{equation} \label{BC}
 \psi(x,y,t)  =  0,  \qquad \om (x,y,t)  =  0 \quad  \mbox{on} \; \partial
D \; ,
\end{equation}
together with an appropriate initial condition,
\begin{equation} \label{IC}
\om (x,y,0)  =  \om_0(x,y) \quad \mbox{on} \;  D \; .
\end{equation}

The global well--posedness (i.e. existence and uniqueness of smooth
solutions) of
the dissipative model (\ref{eqn})--(\ref{IC}) can be obtained similarly as
in, for
example, \cite{Barcilon,Dymnikov,Wu}; see also \cite{Bennett}. Steady wind
forcing
has been used in numerical simulations \cite{Cessi} and Duan  \cite{Duan}
has shown
the existence of time periodic quasigeostrophic response of time periodic
wind forcing by means of a Leray--Schauder topological degree argument and
Browder's
principle.  In this paper it is  assumed that the wind forcing $f(x,y,t)$
is temporally
almost periodic  and a  concept of pullback attraction \cite{CKS,KS2} will be
used to establish the existence of a unique temporally almost periodic
solution of
(\ref{eqn})--(\ref{BC}) under appropriate constraints on the model
parameters. The
main result is

\begin{theorem} \label{mth}
Assume that
$$
\frac{r}2 + \frac{\pi \nu}{|D|} > 	\frac12 \beta \left(\frac{|D|}{\pi}
+1\right)
$$
and that the wind forcing $f(x,y,t)$ is temporally almost periodic with its
$L^2(D)$--norm bounded uniformly  in time $t$ $\in$ $\rit$ by
$$
||f(\cdot,\cdot,t)|| \leq \sqrt{\frac{\pi r}{|D|}} \;\; \left[\frac{r}2 +
\frac{\pi
\nu}{|D|}
	-\frac12 \beta \left(\frac{|D|}{\pi} +1 \right) \right]^{\frac32}.
$$
Then the dissipative quasigeostrophic model (\ref{eqn})--(\ref{BC})
has a  unique temporally almost periodic solution that exists for all time $t$
$\in$ $\rit$.
\end{theorem}
The necessary terminology  will be presented as required in the text and
proof that
follow. Some mathematical preliminaries are stated below, while
dissipativity and
strong contraction properties of QG flow are established in Section 2.
Background
material on pullback attraction for nonautonomous systems is presented in
Section  3
and  that for almost periodicity  in Section  4, where it is  applied
to the QG model under consideration to complete the proof of Theorem
\ref{mth}.

Standard abbreviations $L^2$ $=$ $L^2(D)$, $H^k_0$ $=$ $H^k_0(D)$,
$k = 1, 2,\ldots$,  are used for the common Sobolev spaces in fluid mechanics
\cite{Temam},
with $<\cdot, \cdot>$ and $\| \cdot \|$ denoting the  usual scalar product
and norm,
respectively, in $L^2$. We need the following properties and  
estimates (see also \cite{Dymnikov}) of the
Jacobian
operator $J: H_0^1 \times H_0^1 \rightarrow  L^1$.    
\begin{eqnarray}
\int_D J(f,g) h \,dxdy & = & - \int_D J(f,h) g \,dxdy,
	\label{estimate0}
\\[2 ex]
\int_D J(f,g) g \,dxdy & = & 0,
\label{estimate1}
\\[2 ex]
\left|\int_D J(f,g)  \,dxdy \right| & \leq &  \|\nabla f\| \|\nabla g\|
\label{estimate2}
\end{eqnarray}
for all $f$, $g$, $h$ $\in$ $H^1_0$, and
\begin{eqnarray}
\left|\int_D J(\Delta f, g) \Delta h \,dxdy \right|
 & \leq & \sqrt{\frac{2|D|}{\pi}} \|\Delta f\|  \; \|\Delta g\|  \;\|\Delta h\|
\label{estimate2.5}
\end{eqnarray}
for all $f$, $g$, $h$ $\in$ $H^2_0$, as well as the {\em  Poincar\'e
inequality\/}
\cite{Gilbarg-Trudinger}
\begin{eqnarray}
\|g\|^2 = \int_D g^2(x,y) \,dxdy \leq  \frac{|D|}{\pi} \int_D |\nabla g|^2
\,dxdy
= \frac{|D|}{\pi} \|\nabla g\|^2
\end{eqnarray}
for $g$ $\in$ $H^1_0$, and  {\em Young's inequality \/}
\cite{Gilbarg-Trudinger}
\begin{equation}
AB \leq \frac{\e}2 A^2 + \frac{1}{2 \e}B^2,
\end{equation}
where $A,B$ are non-negative real numbers and $\e >0$. \\

\section{Dynamics of dissipative QG flow}

We first show that the equation (\ref{eqn}) with  boundary conditions
(\ref{BC}) is
a dissipative system in the sense (\cite{Hale,Temam}, see  also
\cite{Duan}) that all
solutions $\om(x,y,t)$ approach a bounded set in 
the phase space $L^2$ as time goes to
infinity
provided that the $L^2$ norm of the forcing term is uniformly bounded in
time and
that the system parameters satisfy the inequality of Theorem \ref{mth}.
Then  we show
that the system is strongly contracting under the  restriction on the
magnitude of the
$L^2$ norm of the forcing term assumed in Theorem \ref{mth}.

\subsection{Dissipativity property}

Define the solution operator $S_{t,t_0}$ $:$ $L^2$ $\to$ $L^2$  by
$S_{t,t_0} \om_0$ $:=$ $\om(t)$ for $t$ $\geq$ $t_0$, where $\om(t)$ is the
solution of the QG equations in $L^2$ starting at $\om_0 $ $\in$ $L^2$
at time $t_0$.  Since the the dissipative QG equations
(\ref{eqn})--(\ref{BC}) are
strictly parabolic, the solution operators $S_{t,t_0}$ exist and are
compact for all
$t$ $>$ $t_0$; see, for example, \cite{Temam}.  In fact, the $S_{t,t_0}$
are compact in
$H^k_0$ for all $k \geq 0$ and so, in particular, $S_{t,t_0}B$ is a compact
subset
of  $L^2$  for each $t$ $>$ $t_0$ and every closed and bounded subset $B$
of $L^2$.\\

Multiplying (\ref{eqn}) by $\om$ and integrating over $D$, we obtain
\begin{eqnarray}
\frac12 \frac{d}{dt}\|\om\|^2  &  = &  -\nu \|\nabla \om\|^2 -
r \|\om\|^2 + \int_D f(x,y,t)\om \,dxdy
 	\label{estimate3}   \\ [2ex]
&  & \quad  - \int_D J(\psi, \om)\om \,dxdy -  \beta \int_D \psi_x \om\,dxdy.
\nonumber
\end{eqnarray}
Now $\int_D J(\psi, \om)\om \,dxdy$ $=$ $0$ by (\ref{estimate0}) and  from
the Young and Poincar\'e inequalities we have
\begin{eqnarray*}
\left|\beta \int_D \psi_x \om \,dxdy \right|
& \leq &  \frac12 \beta \left(\int_D \psi_x^2 \,dxdy + \int_D \om^2
\,dxdy\right)
 \\
& \leq &  \frac12 \beta \left(\frac{|D|}{\pi}\int_D \om^2 \,dxdy + \int_D
\om^2 \,dxdy \right),
\end{eqnarray*}
that is
\begin{equation} 	\label{estimate4}
\left|\beta \int_D \psi_x \om \,dxdy \right| \leq
\frac12 \beta \left(\frac{|D|}{\pi} +1 \right) \|\om\|^2,
\end{equation}
and by the Poincar\'e inequality again we also have
\begin{equation} 	\label{estimate5}
-\nu  \|\nabla \om\|^2  \leq  -\frac{\pi \nu}{|D|} \|\om\|^2.
\end{equation}
Now assume that the square-integral of the wind forcing $f(x,y,t)$
with respect to $(x,y)$ $\in$ $D$ is uniformly bounded in time, i.e.
\begin{equation}
\|f(\cdot,\cdot,t)\| \leq M
\end{equation}
for some  positive constant $M$. (This is  mild assumption because a
temporally
almost periodic function  is bounded in time, see  \cite{Besicovitch} and
later).
Then
\begin{eqnarray*}
\left|\int_D f(x,y,t)\om \,dxdy \right|
& \leq &  \frac1{2r} \int_D f^2(x,y,t) \,dxdy + \frac{r}2 \int_D \om^2 \,dxdy
\\
& \leq &  \frac{M^2}{2r} + \frac{r}2 \|\om\|^2.
\label{estimate6}
\end{eqnarray*}

Putting (\ref{estimate4})--(\ref{estimate6}) into (\ref{estimate3}) we obtain
\begin{eqnarray}
 \frac12 \frac{d}{dt}\|\om\|^2 +  \alpha \|\om\|^2  \leq  \frac{M^2}{2r},
\end{eqnarray}
where
\begin{eqnarray}
\alpha =  \frac{r}2 + \frac{\pi \nu}{|D|}-\frac12 \beta \left(\frac{|D|}{\pi}
+1\right).
\label{alpha}
\end{eqnarray}
Then $\alpha$ $>$ $0$ if we assume that
\begin{eqnarray}
  \frac{r}2 +\frac{\pi \nu}{|D|} > \frac12 \beta \left(\frac{|D|}{\pi}
+1\right),
\end{eqnarray}
which is in fact the first constraint   of Theorem \ref{mth}. Thus, by the
Gronwall
inequality, we have
\begin{eqnarray}
\|\om\|^2 \leq \|\om_0\|^2  e^{-2\alpha t} 	+
\frac{M^2}{2r\alpha} \left(1 - e^{-2\alpha t}\right) .
\end{eqnarray}
Hence all solutions $\om$ enter the closed and bounded set
$${\cal B} = \left\{\om: \; \|\om\|  \leq   \frac{M}{\sqrt{2r\alpha}}  \right\}
$$
in finite time and stay there. The set ${\cal B}$ is thus an absorbing set
of the
system and is positively invariant in the sense that $S_{t,t_0} {\cal B}$
$\subset$
${\cal B}$ for all $t$ $\geq$ $t_0$ and $t_0$ $\in$ $\rit$.\\

For later purposes  note  that the solution operator $S_{t,t_0}$  satisfies
$S_{t_0,t_0}$ $=$ $id.$ and $S_{t_2,t_1}\circ S_{t_1,t_0}$ $=$
$S_{t_2,t_0}$ for any
$t_0$ $\leq$ $t_1$ $\leq$ $t_2$, that is $\{S_{t,t_0}\ : \: t \geq t_0, t_0
\in \rit\}$
is a nonautonomous process or cocycle mapping. In addition, it follows from
existence and uniqueness theory that $(t,t_0,\om_0)$ $\to$ $S_{t,t_0} \om_0$
is continuous. Hence, in particular, when  the forcing  $f$ is independent
of time
there exists a global autonomous attractor defined by
$$
{\cal A}_0 = \bigcap_{t\geq 0} S_{t,0}{\cal B},
$$
which is a nonempty compact subset of $L^2$,  and is invariant under the
autonomous
semigroup $\{S_{t,0} \ : \: t \geq 0\}$ in the sense that $S_{t,0} {\cal
A}_0 $ $=$
${\cal A}_0$ for all $t$ $\geq$ $0$.

\subsection{Strong contraction property}

Now consider two trajectories $\omega^{(i)}$  corresponding to initial values
$\omega_0^{(i)}$ $\in$ ${\cal B}$, $i$ $=$ $1$ and $2$. Note that these
trajectories remain inside ${\cal B}$. Their difference $\delta\omega$ $=$
$\omega^{(1)} - \omega^{(2)}$ satisfies the equation
$$
 \delta\omega_t + J\left(\psi^{(1)}, \om^{(1)}\right) - J\left(\psi^{(2)},
\om^{(2)}
\right) + \beta \delta\psi_x =  \nu \Delta \delta\om -  r\delta\om.
$$
Similarly to the proof  above it can be shown from this equation that
\begin{equation}\label{eqes}
\frac12 \frac{d}{dt} \|\delta\om\|^2 + \int_D \delta J \delta \ \om \,dxdy
 +\beta \int_D \delta \psi_x \, \delta \om  \,dxdy
 =   -\nu \|\nabla \delta \om\|^2 - r \|\delta \om\|^2
\end{equation}
where
$$
\delta J(\psi, \om) := J(\psi^{(1)}, \om^{(1)})  - J(\psi^{(2)}, \om^{(2)}).
$$
Now from the properties (\ref{estimate0})--(\ref{estimate2.5})
of the Jacobian $J$
we have
\begin{eqnarray*}
\left| \int_D \delta J \delta \om \,dxdy \right|
& = &
\left|\int_D \left(J\left(\psi^{(1)},\om^{(1)} \right) -  J\left(\psi^{(2)},
\om^{(2)}\right) \right) \left(\om^{(1)}-\om^{(2)}\right) \,dxdy\right|
\\ [2ex]
& = &  \left|\int_D J\left(\psi^{(1)},\om^{(1)}\right) \om^{(2)} \,dxdy +
\int_D J\left(\psi^{(2)},\om^{(2)}\right) \om^{(1)} \,dxdy \right|
\\[2ex]
& = & \left| \int_D J\left(\psi^{(1)},\om^{(1)}\right) \om^{(2)}\,dxdy -
 \int_D  J\left(\psi^{(2)}, \om^{(1)}\right) \om^{(2)} \,dxdy  \right|
\\[2ex]
& = & \left|\int_D J\left(\psi^{(1)} - \psi^{(2)}, \om^{(1)}\right)
\left(\om^{(1)}
- \om^{(2)}\right)\,dxdy \right|
\\[2ex]
& = & \left|\int_D J\left(\delta \psi, \om^{(1)}\right)\delta \om \,dxdy\right|
\\[2ex]
& = & \left|\int_D J\left( \om^{(1)}, \delta \psi \right)\delta \om \,dxdy\right|
\\[2ex]
& = & \left|\int_D J\left(\Delta \psi^{(1)}, \delta \psi \right)\Delta \delta \psi \,dxdy\right|
\\[2ex]
&\leq &\sqrt{\frac{2|D|}{\pi}} \|\Delta \psi^{(1)}\| \; \|\Delta \delta \psi\| \;\|\Delta \delta \psi\|
\\[2ex]
& =  &  \sqrt{\frac{2|D|}{\pi}} \|\om^{(1)} \|  \; \|\delta \om \|^2 ,
\end{eqnarray*}
where in the last two steps,we have used  (\ref{estimate2.5})
with $f= \psi^{(1)}, g= h=\delta \psi$,  and the fact
 $\Delta \delta \psi$ $=$ $\delta \Delta \psi$ $=$ $\delta \om$.
Using this and  noting that $\om^{(1)}$ is in the positively invariant
absorbing set ${\cal B}$ so $\|\om^{(1)}\|$ $\leq$ $M/\sqrt{2r\alpha}$,
we have
\begin{eqnarray}
\left| \int_D \delta J \delta \om \,dxdy \right|
& \leq  &
 \sqrt{\frac{2|D|}{\pi}} \|\om^{(1)} \|  \; \|\delta \om \|^2   \nonumber
\\[2ex]
& \le &
\sqrt{\frac{2|D|}{\pi}} \frac{M}{\sqrt{2r\alpha}} \|\delta \om\|^2 \nonumber
\\[2ex]
& = &\sqrt{\frac{|D|}{\pi r \alpha}}  \; \;  M    \|\delta \om\|^2.
\label{bigestimate}
\end{eqnarray}
Then from equation (\ref{eqes}), using (\ref{bigestimate})
and (\ref{estimate4}), we obtain
\begin{eqnarray*}
\frac12 \frac{d}{dt} \|\delta\om\|^2
& = &  -\nu   \|\nabla \delta \om\|^2 - r \|\delta \om\|^2
 - \int_D \delta J \delta \om \,dxdy
\\
& & \quad - \beta \int_D \delta \psi_x \delta \om
\,dxdy
\\[2 ex]
& \leq & -\nu   \|\nabla \delta \om\|^2 - r \|\delta \om\|^2
 + \left|\int_D \delta J \delta \om \,dxdy\right|
\\
& & \quad + \left|\beta \int_D \delta \psi_x \delta \om  \,dxdy \right|
\\[2 ex]
& \leq &   -\frac{\pi \nu}{|D|} \|\delta \om\|^2 - r \|\delta \om\|^2
 + \sqrt{\frac{|D|}{\pi r \alpha}}  \; \;  M    \|\delta \om\|^2   
\\
& & \quad + \frac12 \beta \left(\frac{|D|}{\pi} +1 \right) \|\delta \om\|^2
\\[2 ex]
& \leq &   -\gamma  \|\delta \om\|^2
\end{eqnarray*}
where
$$
\gamma :=  r + \frac{\pi \nu}{ |D|} -
\frac12 \beta \left(\frac{|D|}{\pi} +1 \right)
-  \sqrt{\frac{|D|}{\pi r \alpha}}  \; \;  M.
$$
Note that $\gamma > \alpha - \sqrt{\frac{|D|}{\pi r \alpha}}  \; \;  M$.
Thus,  $\gamma > 0$ if we assume that
\begin{eqnarray} \label{parb}
 \| f(\cdot,\cdot,t)\| \leq M < \sqrt{\frac{\pi r}{|D|}} \alpha^{\frac32},
\end{eqnarray}
for all $t$ $\in$ $\rit$.  Here $\alpha$ is defined in (\ref{alpha}), so
(\ref{parb})  holds because of the assumption  on the $L^2$ norm of $f$ in
Theorem \ref{mth}. This gives
$$
\|\delta \om(t)\|^2 \leq \|\om_0\| e^{-2\gamma t} \to 0 \quad \mbox{as} \ \ t
\to \infty,
$$
for solutions starting within the positively invariant absorbing set ${\cal
B}$.
This is the desired strong contractive condition. This means there is  a
unique
solution $\om^{*}(t)$ in ${\cal B}$ to which all other solutions converge.
This
solution $\om^{*}(t)$ can be determined by the pullback convergence to be
discussed in the following two Sections.

\section{Nonautonomous dynamical systems}

In order to show existence of temporally almost periodic solutions,
we need some results from the theory of nonautonomous dynamical systems.
Consider first an autonomous dynamical system on a metric space $P$
described by
a group $\theta$ $=$ $\{\theta_t\}_{t \in \rit}$ of mappings of $P$ into
itself.

Let $X$ be a complete metric space and consider a continuous mapping
$$
 \Phi : \rit^{+} \times P \times X \to   X
$$
satisfying the properties
$$
\Phi(0,p,\cdot) = {\rm id}_X, \qquad
\Phi(\tau +t,p,x)  =  \Phi(\tau,\theta_t  p, \Phi(t,p,x))
$$
for all $t$, $\tau$ $\in$ $\rit^{+}$, $p$ $\in$ $P$ and $x$ $\in$ $X$.
The mapping $\Phi$ is called a cocycle on $X$ with respect to $\theta$ on $P$.

The appropriate concept of an attractor for a nonautonomous cocyle systems is
the {\em pullback attractor\/}. In contrast to autonomous attractors it
consists of a family subsets of the original state space $X$ that are
indexed by
the cocycle parameter set.
\begin{definition}\label{pba}
A family $\widehat{A}$ $=$ $\{A_p\}_{p \in P}$ of nonempty compact sets of
$X$ is
called a {\rm pullback attractor\/} of the cocycle $\Phi$ on $X$ with
respect to
$\theta_t$ on $P$ if it is ${\Phi}$--invariant, i.e.
$$
\Phi(t,p,A_p) = A_{\theta_t} p  \qquad \mbox{for all} \quad t \in \rit^{+},
p \in P,
$$
and {\rm pullback attracting}, i.e.
$$
\lim_{t \to \infty} H^{*}_X\left(\Phi(t,\theta_{-t}p,D), A_p\right) = 0
\qquad \mbox{for all} \quad D \in K(X), \  p \in P,
$$
where $K(X)$ is the space of all nonempty compact subsets of the metric
space $(X,d_X)$.
\end{definition}
Here $H^{*}_X$ is the Hausdorff semi--metric between nonempty compact subsets
of $X$, i.e. $H^{*}_X(A,B)$ $:=$ $\max_{a \in A} {\rm dist}(a,B)$ $=$
$\max_{a\in A} \min_{b\in B} d_X(A,b)$ for $A$, $B$ $\in$ $K(X)$.

The following theorem combines  several known results.  See Crauel and
Flandoli
\cite{CF}, Flandoli and Schmalfu{\ss} \cite{FS1}, and Cheban \cite{C1} as
well as
\cite{CKS,KS2}  for this and various related proofs.
\begin{theorem} \label{th1}
Let $\Phi$ be a continuous cocycle on a metric space $X$ with respect to a
group
$\theta$ of continuous mappings on a metric space $P$. In addition, suppose
that
there is a nonempty compact subset $B$ of $X$ and that for every $D$ $\in$
$K(X)$
there exists a $T(D)$ $\in$ $\rit^{+}$, which is independent of $p$ $\in$ $P$,
 such that
\begin{equation}\label{fa}
\Phi(t,p,D) \subset B \quad \mbox{for all} \quad t > T(D).
\end{equation}
Then  there exists a unique pullback attractor
$\widehat{A}$ $=$
$\{A_p\}_{p \in P}$ of the cocycle $\Phi$ on $X$, where
\begin{equation}\label{pbat}
A_p = \bigcap_{\tau \in \rit^{+}} \overline{\bigcup_{t > \tau \atop t \in
\rit^{+}}
\Phi\left(t,\theta_{-t}p,B\right)}.
\end{equation}
Moreover, the mapping $p$ $\mapsto$ $A_p$  is upper semicontinuous.
\end{theorem}
Moreover, in \cite{CKS} it is shown that the  pullback attractor consists
of a single
trajectory when the cocycle dynamics are in fact strongly contracting.
\begin{theorem}\label{th2}
Suppose that the cocycle $\Phi$ in Theorem \ref{th1} is strongly
contracting inside
the absorbing set $B$. Then the pullback attractor consists of  singleton
valued
sets, i.e.   $A_p$ $=$ $\{a^*(p)\}$,  and the mapping $p$ $\mapsto$
$a^*(p)$ is
continuous.
\end{theorem}

\section{Almost periodicity}

A function $\varphi$  $:$ $\rit$ $\to$ $X$, where $(X,d_X)$ is  a
metric space, is called {\em almost periodic\/} \cite{Besicovitch}
 if for every $\varepsilon$ $>$ $0$  there exists a
relatively dense subset $M_{\varepsilon}$ of $\rit$ such that
$$
d_X \left(\varphi (t+ \tau ), \varphi (t) \right) <  \varepsilon
$$
for all $t$ $\in$ $\rit$ and $\tau$ $\in M_{\varepsilon }$.  A subset $M$
$\subseteq$ $\rit$ is called {\em relatively dense} in $\rit$ if there exists
a positive number $l$ $\in$ $\rit$ such that for every $a$ $\in$ $\rit$ the
interval $[a,a+l]\bigcap \rit$ of length $l$ contains an element of $M$, i.e.
$M\bigcap [a,a+l]$ $\ne$ $\emptyset$ for every $a$ $\in$ $\rit$.

The QG solution operators $S_{t,t_0}$ form a cocycle mapping on $X$ $=$
$L^2$ with
parameter set $P$ $=$ $\rit$, where $p$ $=$ $t_0$, the initial time, and
$\theta_t t_0$ $=$ $t_0+t$, the left shift by time $t$. However, the space
$P$ $=$
$\rit$ is not compact here. Though more complicated, it is more useful to
consider
$P$ to be the closure of the subset $\{\theta_t f, t \in \rit\}$, i.e. the
hull of $f$,  in
the metric  space $L^2_{loc}\left(\rit,L^2(D)\right)$ of locally
$L^2(\rit)$--functions
$f$ $:$ $\rit$ $\to$ $L^2(D)$ with the metric
$$
d_P(f,g) :=  \sum_{N=1}^{\infty} 2^{-N} \min\left\{1, \sqrt{\int_{-N}^N
\|f(t)-g(t)\|^2 \, dt} \right\}
$$
with  $\theta_t$ defined to be  the left shift operator, i.e. $\theta_t
f(\cdot)$ $:=$
$f(\cdot+t)$.  By a classical result \cite{Besicovitch,Sell},  a function
$f$ in the
above metric space is almost periodic if and only if the the  hull of $f$
is compact
and minimal. Here minimal means nonempty, closed and invariant with respect
to the
autonomous dynamical system generated by the shift operators $\theta_t$
such that
with no proper subset has these properties. The cocycle mapping is defined
to be the
QG solution $\om(t)$ starting at $\om_0$ at time $t_0$ $=$ $0$ for a given
forcing
mapping $f$ $\in$ $P$, i.e.
$$
\Phi(t,f,\om_0) := S_{t,0}^f \ \om_0,
$$
where we have included a superscript $f$ on $S$ to denote the dependence on
the forcing
term $f$. (This dependence is in fact continuous). The cocycle property
here follows
from the fact  that $S_{t,t_0}^f \om_0$ $=$ $S_{t-t_0,0}^{\theta_{t_0}f}\
\om_0$ for
all $t$ $\geq$ $t_0$, $t_0$ $\in$ $\rit$, $\om_0$ $\in$ $L^2$ and $f$ $\in$
$P$.

\begin{theorem}\label{th3}
Let the assumptions of Theorem  \ref{mth} hold. Then  the dissipative QG
model
(\ref{eqn})--(\ref{BC})  has a unique  almost periodic solution $\om^{*}$
defined by
$$
\om^{*}(t) := a^*\left(\theta_t f\right),  \qquad t \in \rit,
$$
where $\{a^{*}(p)\}$ is  the  singleton valued pullback
attractor--trajectory of the
cocycle $\Phi(t,f,\om_0)$ on $L_2(D)$,  $P$ is the hull in the metric space
$L^2_{loc}\left(\rit,L^2(D)\right)$ of  the almost periodic forcing term
$f$ and the
$\theta_t$  are the left shift operators on  $P$.
\end{theorem}
This is proved as follows. By Theorems \ref{th1} and \ref{th2} the pullback
attractor
exists, consists of  singleton valued components $\{a^*(p)\}$ and the mapping
$p$ $\mapsto$ $a^{*}(p)$ is continuous on $P$. In fact, the mapping  $p$
$\mapsto$
$a^{*}(p)$ is uniformly continuous on $P$ because $P$ is compact subset of
$L^2_{loc}\left(\rit,L^2(D)\right)$ due to the assumed  almost periodicity.
That is, for every $\varepsilon$ $>$ $0$ there exists a $\delta(\varepsilon)$
$>$ $0$ such that $\|a^*(p)-a^*(q) \|$ $<$ $\varepsilon$ whenever
$d_P(p,q)$ $<$
$\delta$.  Now let  the point $\bar{p}$ ($=$ $f$, the given temporal
forcing function)
be almost periodic and   for $\delta$ $=$ $\delta(\varepsilon)$
$>$ $0$ denote by $M_{\delta}$ the relatively  dense subset of $\rit$ such that
$d_P(\theta_{t+\tau}\bar{p},\theta_t \bar{p})$ $<$ $\delta$ for all $\tau$
$\in$ $M_{\delta}$ and $t$ $\in$ $\rit$. From this and the uniform
continuity we have
$$
\|a^*(\theta_{t+\tau} \bar{p}) - a^*(\theta_t \bar{p})\| < \varepsilon
$$
for all $t$ $\in$ $\rit$ and $\tau$ $\in$ $M_{\delta(\varepsilon)}$. Hence
$t$ $\mapsto$ $\om^*(t)$ $:=$  $a^*(\theta_{t}\bar{p})$ is almost periodic.
It is unique as the single-trajectory pullback attractor is the only trajectory
that exists and is bounded for the entire time line.

This completes the proof of the main result,  Theorem \ref{mth}.

\end{document}